\documentclass[letterpaper,12pt]{article}
\usepackage{tabularx} % extra features for tabular environment
\usepackage{amsmath}  % improve math presentation
\usepackage[dvipdfmx]{graphicx} % takes care of graphic including machinery
\usepackage[margin=1in,letterpaper]{geometry} % decreases margins
\usepackage{cite} % takes care of citations
\usepackage[final]{hyperref} % adds hyper links inside the generated pdf file
\usepackage{authblk}
\hypersetup{
	colorlinks=true,       % false: boxed links; true: colored links
	linkcolor=blue,        % color of internal links
	citecolor=blue,        % color of links to bibliography
	filecolor=magenta,     % color of file links
	urlcolor=blue         
}

\newcommand{\Tabref}[1]{Table \ref{#1}}
\newcommand{\Figref}[1]{Figure \ref{#1}}
\newcommand{\Equref}[1]{Equation \ref{#1}}
\begin{document}

\title{Study of the $h\gamma Z$ coupling using $e^+e^-\xrightarrow{}h \gamma$ at the ILC}
\author{Yumi Aoki$^{1}$, Keisuke Fujii$^{2}$, Sunghoon Jung$^{3}$, Junghwan Lee$^{3}$, Junping Tian$^{4}$, Hiroshi Yokoya$^{5}$
\\
on behalf of the ILD concept group}
\affil{SOKENDAI$^{1}$, KEK$^{2}$, Seoul National University$^{3}$,University of Tokyo$^{4}$, KIAS$^{5}$}

\maketitle

\begin{abstract}
 We study the $e^+e^-\to h \gamma$ process at the International Linear Collider (ILC) to probe new physics in the $h\gamma Z$ coupling. The study is performed at a center of  mass energy of 250 GeV and is based on the full simulation of the International Large Detector (ILD). The expected signal significance is found to be 0.53 $\sigma$ for an integrated luminosity of 2000 fb$^{-1}$ in the case of the standard model. The corresponding 95 $\%$ confidence level upper limit for the signal cross section is 1.08 fb for left-handed beam polarization. 
 %This study was performed in the framework of the ILD concept.
\footnote{
Talk presented at the International Workshop on Future Linear Colliders (LCWS2018), Arlington, Texas, 22-26 October 2018. C18-10-22.}

\end{abstract}

\section{Introduction}

The discovery of the Higgs boson at the Large Hadron Collider (LHC) has completed the standard model particle spectrum. The most important task is now to find physics beyond the standard model. Precision study of the Higgs boson is a powerful tool for this purpose. The International Linear Collider (ILC)\,\cite{Aus0} is an ideal machine to carry out the precision Higgs measurements. 

\begin{figure}[ht] 
        
        \centering \includegraphics[width=0.4\columnwidth]{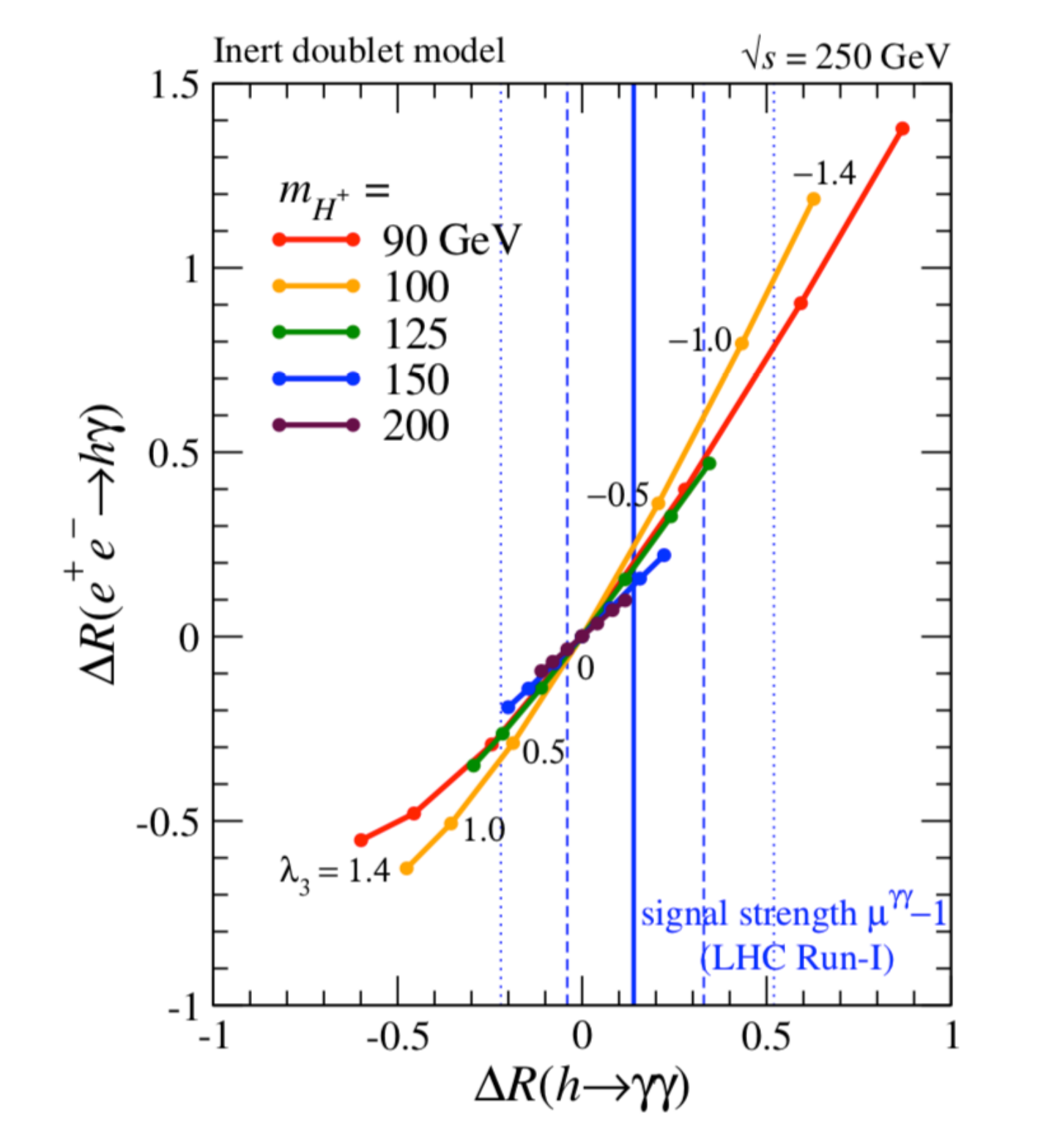}
        \caption{
                \label{fig:1} 
               The relative deviations of the $e^+ e^- \to h\gamma$ cross section and the $h \xrightarrow{} \gamma\gamma$ branching ratio with respect to the Standard Model values~\cite{Aus2}}
\end{figure}

The motivation of our study is to find new physics effects in $h\gamma\gamma$ and $h \gamma Z$ couplings. Since these couplings appear only at the loop level in the standard model, they are potentially very sensitive to new physics and being studied at the LHC. As one example, the expected deviations on the $e^+e^- \to h \gamma$ cross section and the $h \to \gamma \gamma$ branching ratio in the Inert Doublet Model~\cite{Aus2} are shown in \Figref{fig:1}, which suggests that depending on model parameters the deviations can be as large as 100$\%$. 

\begin{figure}[ht] 
        \centering \includegraphics[width=0.4\columnwidth]{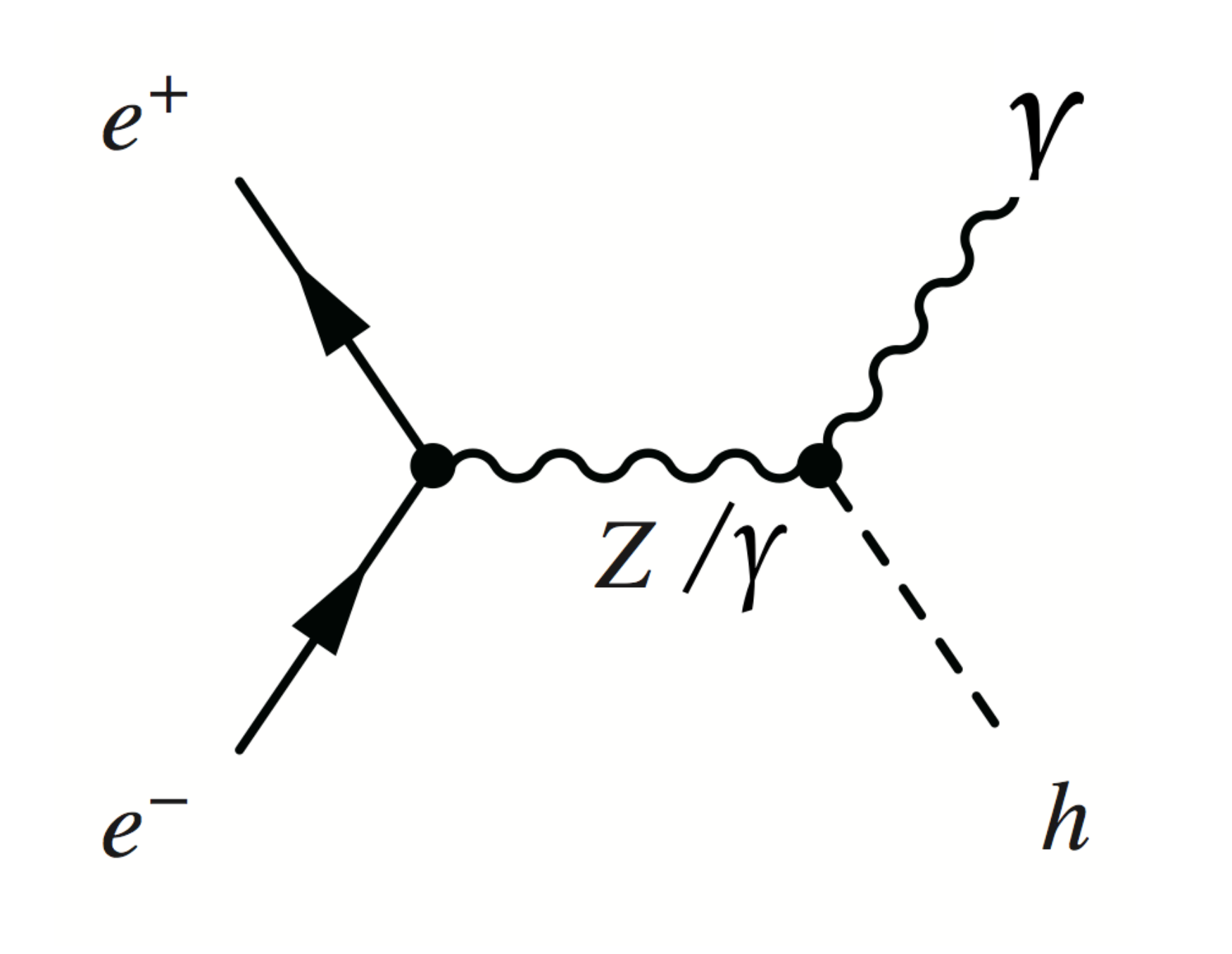}
        \caption{
                \label{fig:6} % spaces are big no-no withing labels
                % things like fig: are optional in the label but it helps
                % to orient yourself when you have multiple figures,
                % equations and tables
                The Feynman diagram of $e^+ e^- \to h \gamma$
        }
\end{figure}

A usual method to measure $h\gamma\gamma$ and $h \gamma Z$ couplings is to use decay branching ratios of $h \to \gamma \gamma / \gamma Z$. It is, however, very challenging to measure the  $h \to  \gamma Z$ branching ratio even at the HL-LHC: only a 3$\sigma$ significance is expected. As a complementary method we study these couplings in a production process at the ILC, $e^+e^- \to h \gamma$ (see \Figref{fig:6}).

  In addition, the $h \gamma Z$ coupling appears in an s-channel photon exchange diagram for the leading single Higgs production process: $e^+ e^- \to h Z$ in the effective field theory framework. It is hence necessary to know the contribution from this diagram. 
  Furthermore, it turns out that the anomalous $h \gamma Z$, $h \gamma \gamma$, $hZZ$, and $hWW$ couplings come from a common set of a  few dimension-6 operators, hence the measurement of the $h \gamma Z$ coupling using $e^+e^- \to h \gamma$ has a potential to provide one very useful constraint on those operators. 
  
%, for example some new heavy particles contributing to the loop,
  
In section 2, we introduce our theoretical framework. In section 3, we describe our experimental method and simulation framework. In section 4, we present our event selection and analysis result. Section 5 gives our plan for further study.

        % read manual to see what [ht] means and for other possible 

\section{Theoretical Framework}
\begin{figure}[ht] 
        
        \centering \includegraphics[width=0.9\columnwidth]{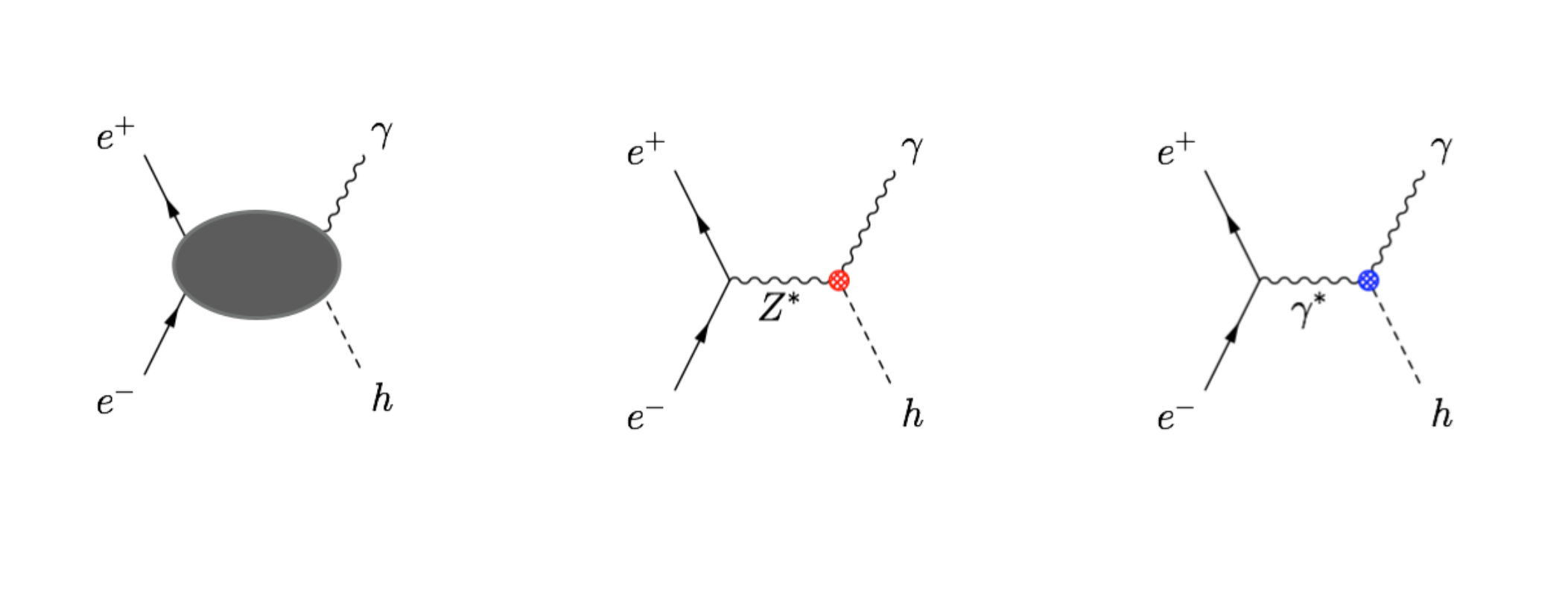}
        \caption{
                \label{fig:5} 
              Diagrams arising from each of the three terms of \Equref{Equ:0}, respectively. }
\end{figure}

\begin{figure}[ht] 
        
        \centering \includegraphics[width=0.9\columnwidth]{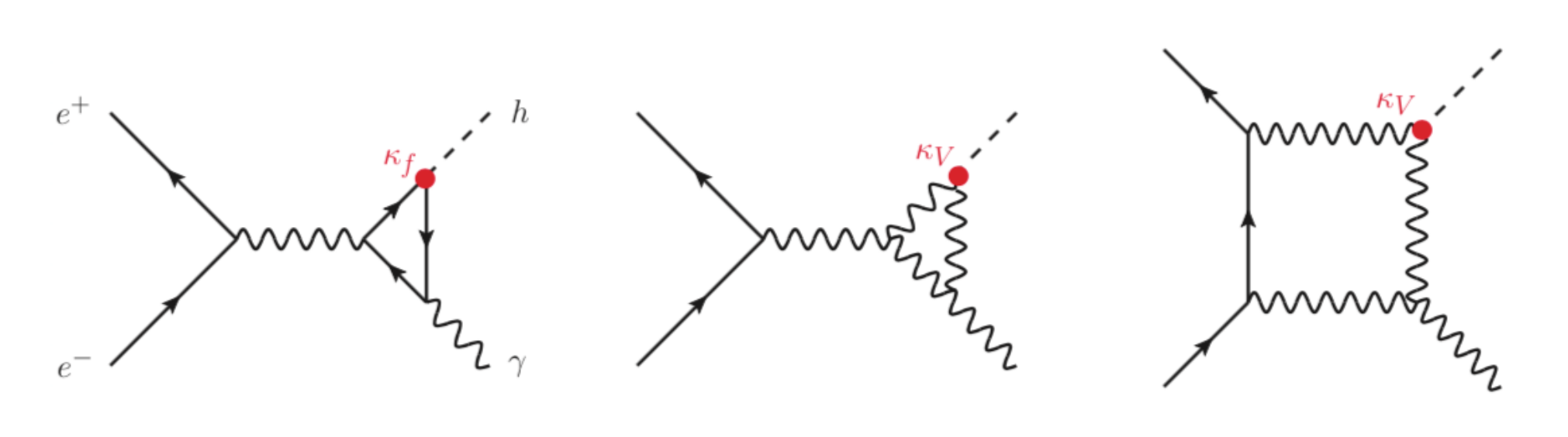}
        \caption{
                \label{fig:7} 
              The loop induced Feynman diagrams in the Standard Model for $e^+e^- \to h \gamma$~\cite{Aus2}}
\end{figure}

In this analysis, we use the effective Lagrangian shown in \Equref{Equ:0} to include new physics contributions to the $e^+ e^- \xrightarrow{} h \gamma $ cross section model-independently,
\begin{eqnarray}
{\cal{L}} _ {h \gamma  } = {\cal{L}} _ { \mathrm { SM } } + \frac { \zeta _ { A Z } } { v } A _ { \mu \nu } Z ^ { \mu \nu } h + \frac { \zeta _ { A } } { 2 v } A _ { \mu \nu } A ^ { \mu \nu } h, 
\label{Equ:0}
\end{eqnarray}
where $\zeta_{AZ}$ and  $\zeta_A$ terms represent respectively effective $h \gamma Z $ and $h \gamma \gamma $ couplings from new physics. $A_{\mu \nu}$, and $ Z_{\mu \nu}$ are field strength tensors. $v$ is the vacuum expectation value. The first term is the Standard Model Lagrangian. The three terms contribute to $e^+e^- \to h \gamma$ process via the Feynman diagrams shown in \Figref{fig:5}, where the first SM diagram represents several loop induced diagrams as shown in \Figref{fig:7}.

The cross section normalized to SM can be written as \Equref{Equ:1} and \Equref{Equ:2}~\cite{Aus3}.
The SM cross sections at $\sqrt{s}$  = 250 GeV are calculated as shown in \Tabref{tbl:1}. The cross sections including effective $h \gamma Z / h\gamma \gamma$ couplings from new physics are calculated as in $P(e^-,e^+)=(+100\%,-100\%)$) , up to interference term. 
\begin{eqnarray}
\frac { \sigma _ { \gamma H } } { \sigma _ { S M } } = 1 - 201 \zeta _ { A } - 273 \zeta _ { A Z }
\label{Equ:1}
\end{eqnarray}

\begin{eqnarray}
\frac { \sigma _ { \gamma H } } { \sigma _ { S M } } = 1 + 492 \zeta _ { A } - 311 \zeta _ { A Z }
\label{Equ:2}
\end{eqnarray}

\begin{table}[htbp]
\begin{center}
\caption{SM cross sections for different beam polarizations ($\sqrt{s}$  = 250 GeV )}
\label{tbl:1} % spaces are big no-no withing labels
\begin{tabular}{|c|c|c|c|} 
\hline
\multicolumn{1}{|c}{$P_{e^-}$ } & \multicolumn{1}{|c|}{$P_{e^+}$} & \multicolumn{1}{c|}{$\sigma_{SM}$[fb]}  \\
\hline
-100$\%$ & +100$\%$ &  0.35\\
+100$\%$ & -100$\%$	& 0.016\\
-80$\%$ & +30$\%$ & 0.20\\

\hline
\end{tabular}
\end{center}
\end{table}

\section{Experimental Method and Simulation Framework}
\subsection{Experimental Method}
In order to determine both $\zeta_{AZ}$ and $\zeta_{A}$, we need two measurements. There are two strategies:\\
1. to measure the cross sections of $e^+ e^- \xrightarrow{} \gamma h$ for two different beam polarizations, and\\
2. to use the measurement of the $h \xrightarrow{} \gamma\gamma$ branching ratio at the LHC to constrain $\zeta_A$ and determine $\zeta_{AZ}$ by just measuring $e^+e^- \to h \gamma$ cross section for one single polarization.\\

\subsection{Simulation framework}

We use fully-simulated Monte-Carlo (MC) samples produced with the ILD detector model~\cite{Aus4}. For event generation, we use Physsim~\cite{Aus7} for signal, and Whizard~\cite{Aus11} for background processes. For detector simulation, we use Mokka~\cite{Aus5}, which is based on Geant4~\cite{Aus12}, and for event reconstruction, we use Marlin in iLCSoft~\cite{Aus6}, where particle flow analysis (PFA) is done with PandoraPFA~\cite{Aus13} and flavor tagging is done with LCFI+~\cite{Aus10}.  The analysis is carried out at $\sqrt{s}$=250 GeV, assuming an integrated luminosity of 2000 fb$^{-1}$.

\section{Event Selection and Results}
\subsection{Event Selection}
The signal channel studied in this paper is $e^+ e^- \to h \gamma $, followed by $h \to b \bar{b}$. In the final states of the signal events, we expect one isolated monochromatic photon with an energy of $E_{\gamma}=\sqrt{s}/{2}\left( 1- {\left(m_h/\sqrt{s}\right)}^2 \right) = 93$~GeV, where $m_h$ is the Higgs mass, and two $b$ jets with an invariant mass consistent with the Higgs mass. The enegy resolution of the electromagnetic calorimeter is given by $\sigma_E=0.16 \times \sqrt{E}$~(GeV), where the photon energy $E$ is in units of GeV~\cite{Aus4}. The energy resolution for the isolated photon is thus 1.5~GeV.
The main background would be $e^+e^-\to \gamma q\bar{q}$, dominantly coming from $e^+e^- \to \gamma Z$.

As pre-selection, we start with identifying one isolated photon with an energy greater than 50 GeV. Sometimes, the reconstruction software PandoraPFA splits calorimetric clusters created by a single high energy photon into several objects. Such split clusters fall within a narrow cone ($\cos \theta_{cone}$=0.998, where $\cos  \theta_{cone}$ is cone angle), and are considered as a single object in the following analysis.The particles other than the photon are clustered into two jets using the Durham algorithm~\cite{Aus15}.

\begin{figure}[ht] 
        
        \centering \includegraphics[width=0.6\columnwidth]{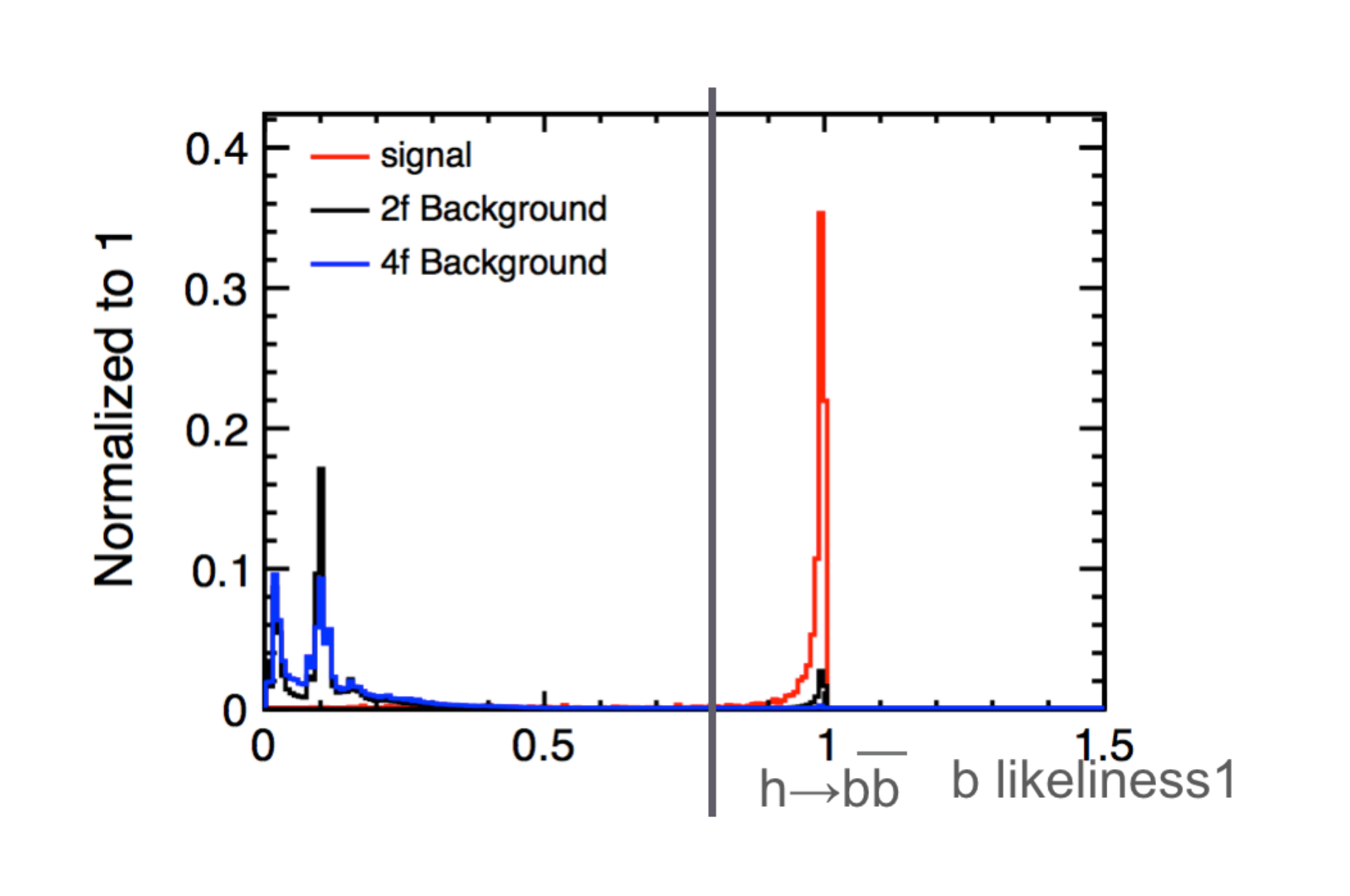}
        \caption{
                \label{fig:8} 
               The distribution of $b$-likeliness for signal and background events with unit normalization}
\end{figure}

In the final selection, the first cut requires two $b$ jets. \Figref{fig:8} shows the distribution normalized to unity of the larger $b$-likeliness value among the two jets for signal and background events. We require this larger $b$-likeliness to be greater than 0.77 to suppress the light flavor $\gamma q\bar{q} $ events. The cut value is optimized to maximize the signal significance defined as
\begin{eqnarray}
\text {significance} = \frac { N _ { S } } { \sqrt { N _ { S } + N _ { B } } },
\label{Equ:7}
\end{eqnarray}
where $N_S$ and $N_B$ are the numbers of signal and background events, respectively.
%We decide the value 0.77 by this graph. 

\if 0
\begin{figure}[ht] 
        
        \centering \includegraphics[width=0.5\columnwidth]{htobb2.pdf}
        \caption{
                \label{fig:9} 
               }
\end{figure}
\fi

\begin{figure}[ht] 
        
        \centering \includegraphics[width=0.6\columnwidth]{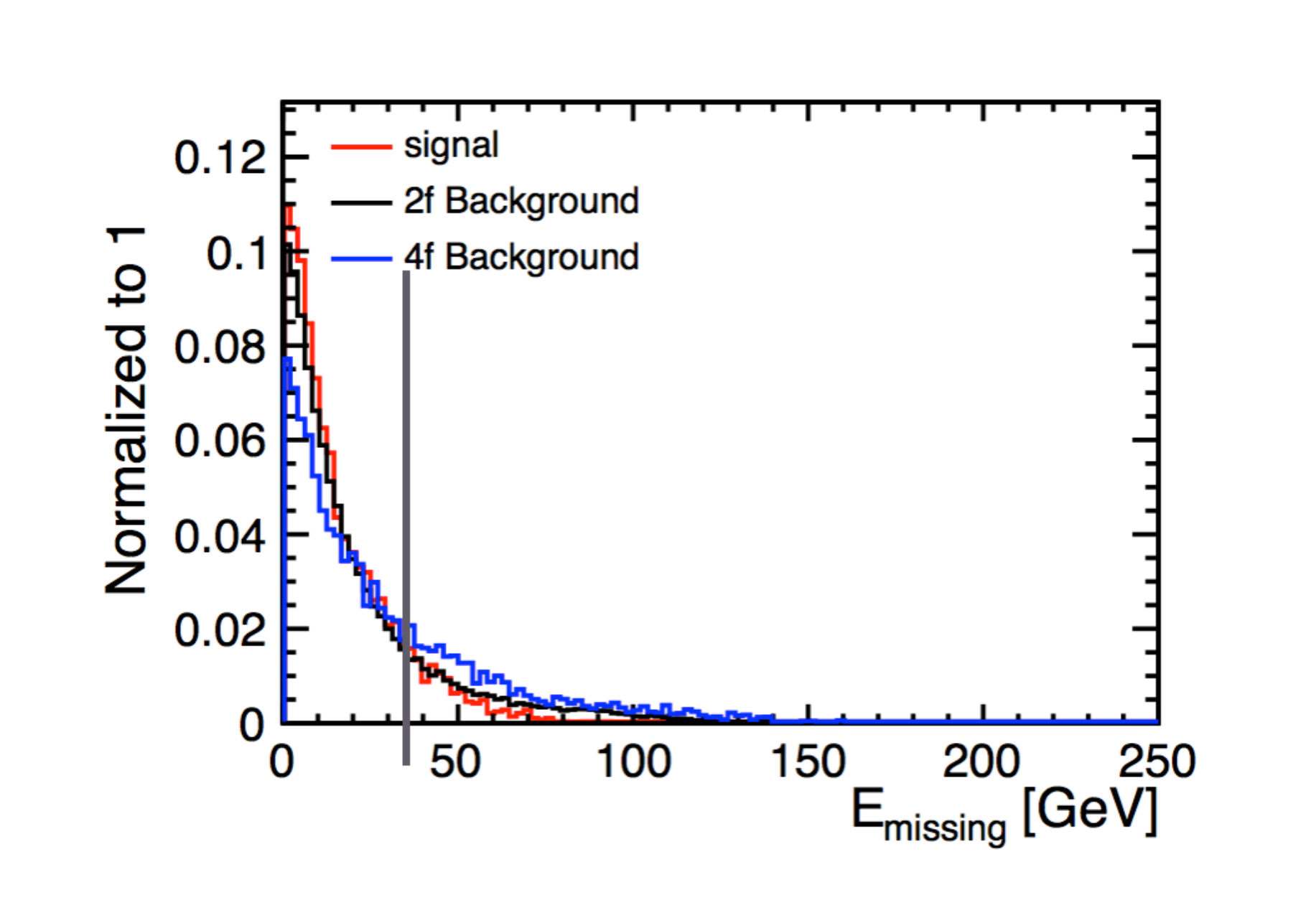}
        \caption{
                \label{fig:10} 
              Normalized missing energy distribution for signal and background events}
\end{figure}

The second cut requires small enough missing energy. \Figref{fig:10} shows the missing energy distribution. We cut the missing energy at 35~GeV. 

\if 0
\begin{figure}[ht] 
        
        \centering \includegraphics[width=0.4\columnwidth]{emis2}
        \caption{
                \label{fig:11} 
               }
\end{figure}
\fi

\begin{figure}[h] 
        
        \centering \includegraphics[width=0.7\columnwidth]{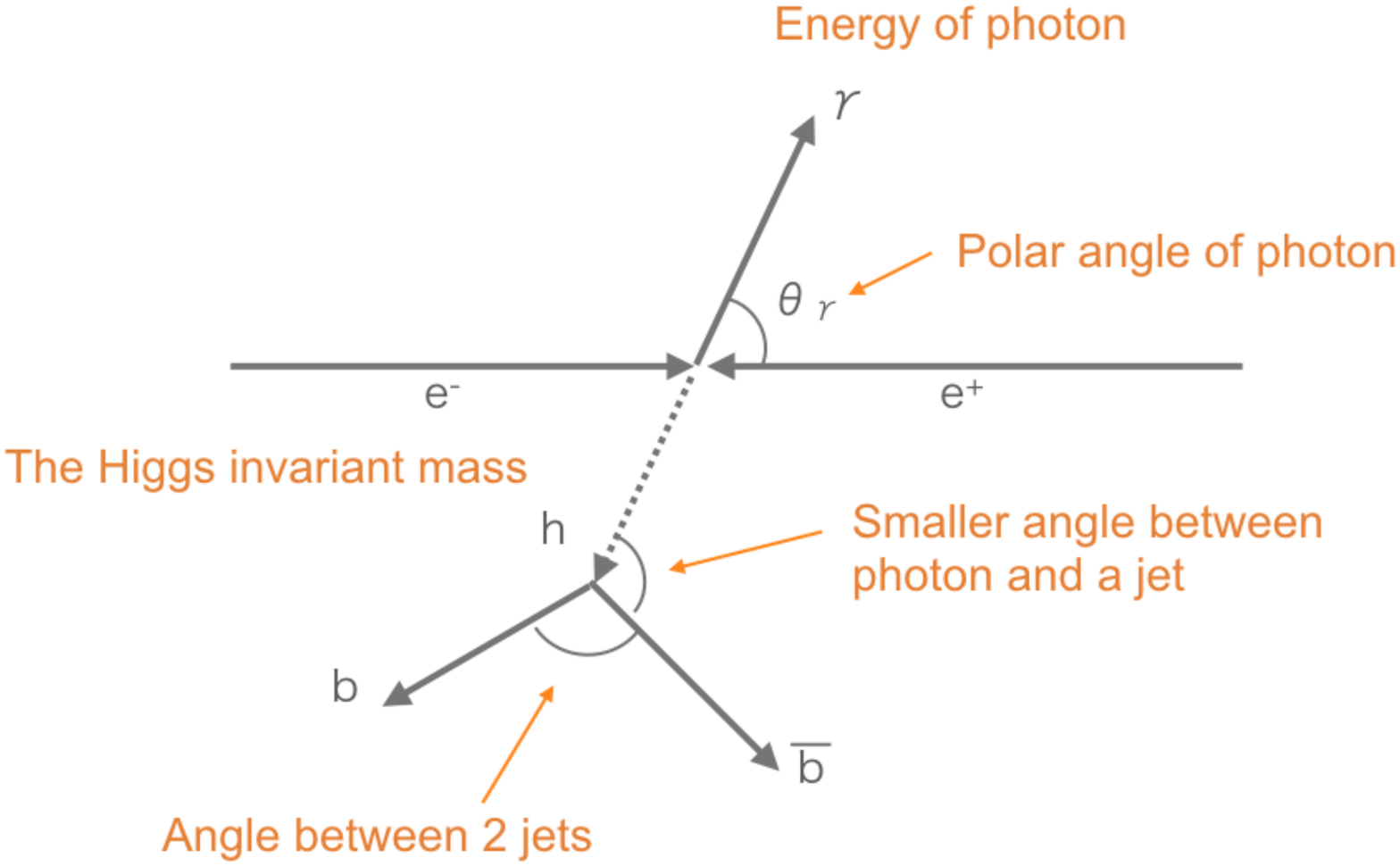}
        \caption{
                \label{fig:12} 
               Input variables for TMVA}
\end{figure}

\begin{figure}[h] 
        
        \centering \includegraphics[width=0.7\columnwidth]{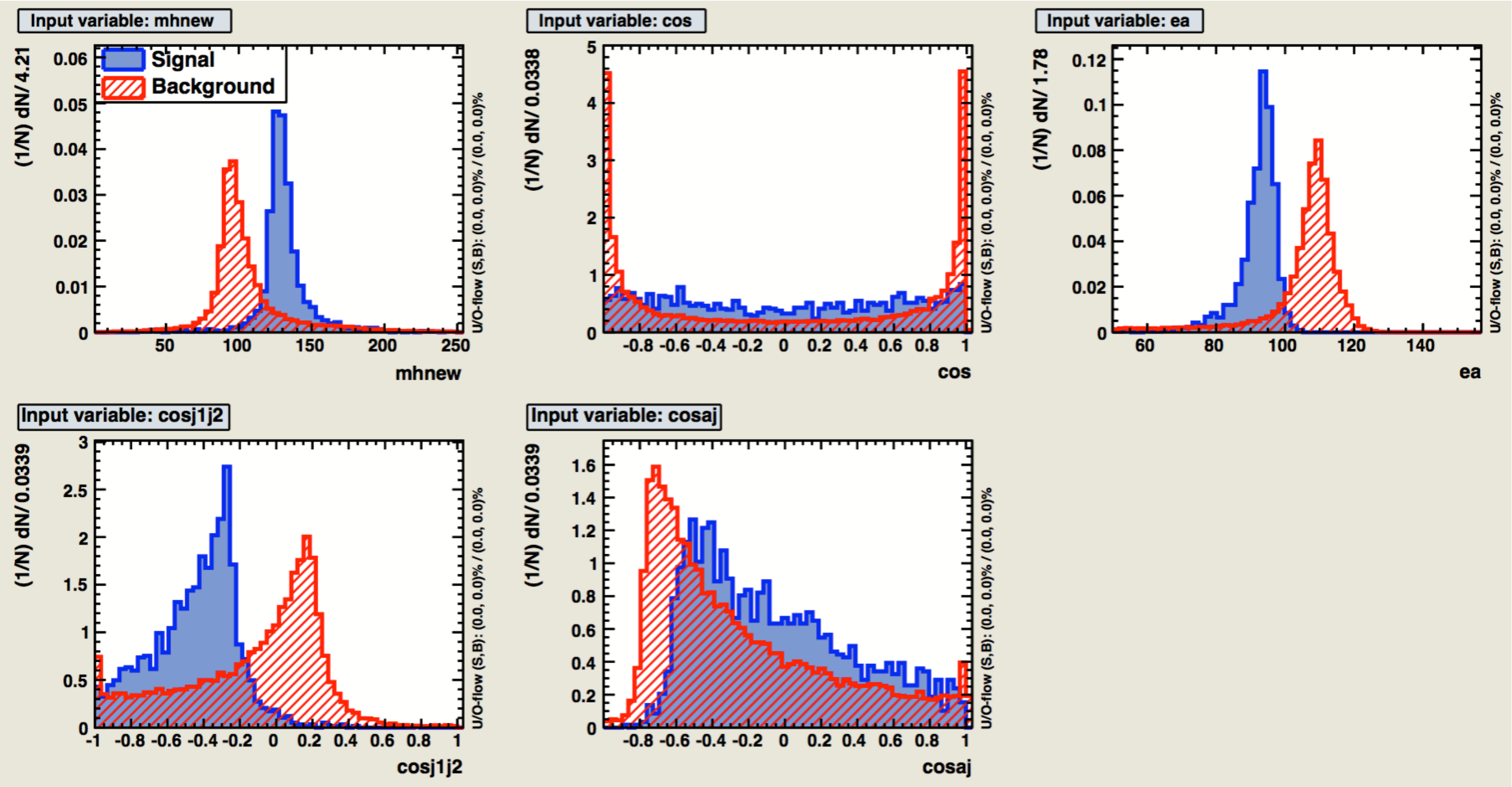}
        \caption{
                \label{fig:14} 
              Distributions of TMVA input variables for signal and background events. }
\end{figure}

To finalize our event selection, we use a multivariate analysis method, the BDT algorithm as implemented in the TMVA package~\cite{Aus8}. It is trained using five input variables: the 2-jet invariant mass, the energy of the isolated photon, its polar angle, the smaller angle between the photon and a jet, and the angle between the two jets. \Figref{fig:12} illustrates these input variables. \Figref{fig:14} shows the distributions of each input variable for signal and background events. The blue histograms are for signal events, and the red histograms are for background events. Our final cut requires the BDT output to be greater than 0.0126.

\begin{figure}[h] 
        
        \centering \includegraphics[width=0.7\columnwidth]{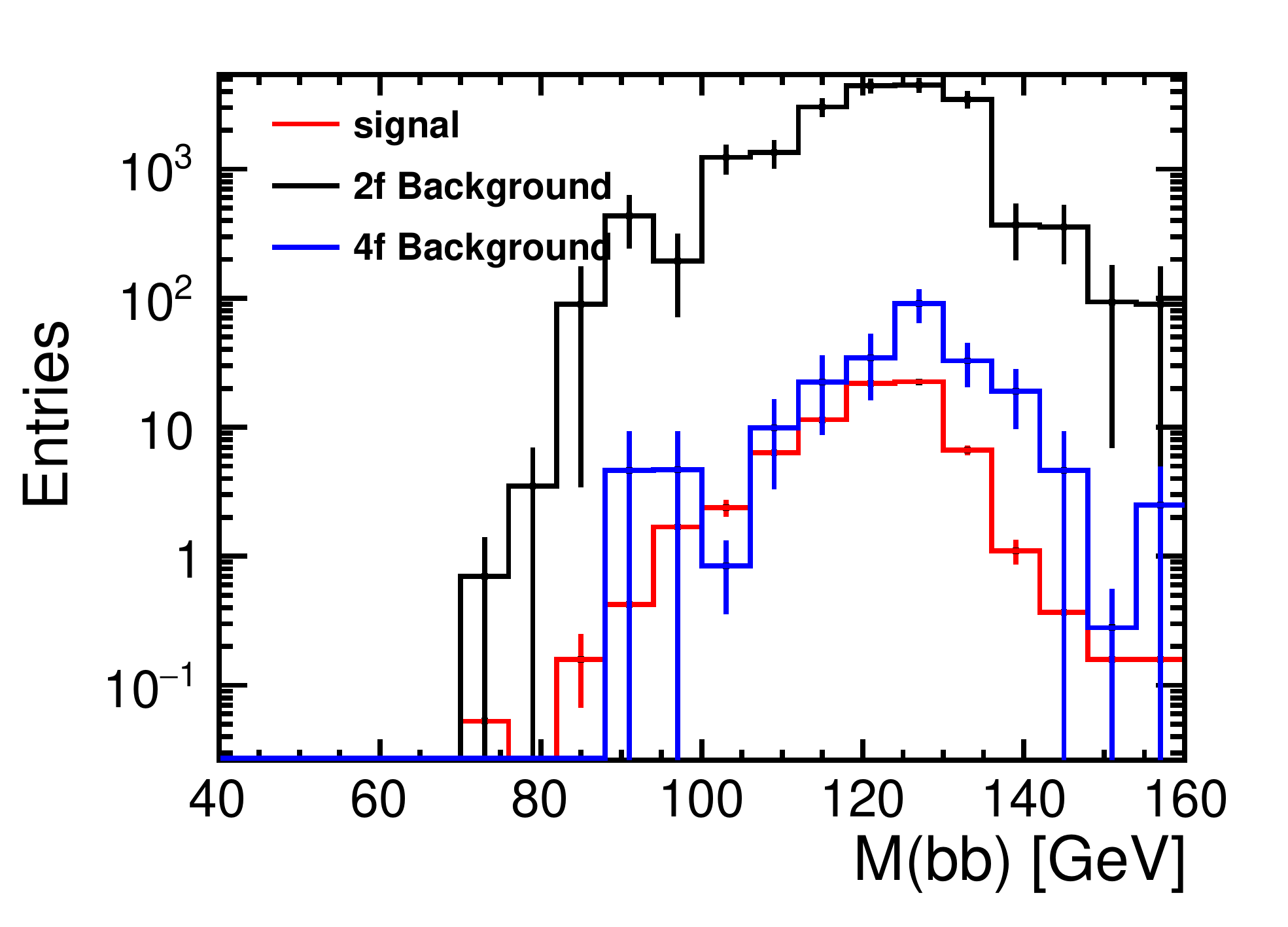}
        \caption{
                \label{fig:15} 
               The distribution of $m(b\bar{b})$ after all the other cuts for the signal and background events normalized to 2000 fb$^{-1}$}
\end{figure}

\Figref{fig:15} shows the distribution of $m(b\bar{b})$ after all the other cuts for the signal and background events normalized to an integrated luminosity of 2000 fb$^{-1}$ for the left-handed beam polarization. The remaining background events are dominated by 2-fermion processes. 

\clearpage
\subsection{Result}
\Tabref{tbl:bins} gives the number of signal and background events, as well as the signal significance after each cut. The significance is defined by \Equref{Equ:7}. 
After all the cuts, the signal significance is expected to be 0.53$\sigma$, for the SM signal process $e^+e^-\to h \gamma$ followed by $h\to b\bar{b}$ decay. 
The 95 $\%$ confidence level upper limit for the cross section at $\sqrt{s}=250$ of $e^+ e^- \to h \gamma$ is calculated using \Equref{Equ:4} to be $\sigma_{h\gamma}^{CL95} < 1.08$ fb, for 2000 fb$^{-1}$ GeV and left handed beam polarizations.

\begin{table}[htbp]
\begin{center}
\caption{The cut table}
\label{tbl:bins} % spaces are big no-no withing labels
\begin{tabular}{|c|c|c|c|} 
\hline
\multicolumn{1}{|c}{ } & \multicolumn{1}{|c|}{Signal} & \multicolumn{1}{c|}{background} & \multicolumn{1}{c|}{Significance} \\
\hline
Expected &237 &   3.14$\times 10^8$ & 0.01\\
Pre selection &222	& 6.54$\times10^7$	&0.02 \\
$b_{tag}\geq$0.8 &200	& 4.96$\times10^6$	&0.09\\
$E_{mis}\leq$35 & 182	& 4.30$\times10^6$	&0.09 \\
$mvabdt \geq$ 0.0126 &75	&1.98$\times10^4$	&0.53\\
\hline
\end{tabular}
\end{center}
\end{table}

\begin{eqnarray}
\sigma = \frac { 1.64 } { \text { significance } } \sigma _ { S M }
\label{Equ:4}
\end{eqnarray}

From this upper limit and \Equref{Equ:1} and \Tabref{tbl:1}, we have 

\begin{eqnarray}
3.09 > \frac { \sigma _ { \gamma H } } { \sigma _ { S M } } = 1 - 201 \zeta _ { A } - 273 \zeta _ { A Z } > 0
~(\mbox{assuming}~ \zeta _ { A } = 0).
\label{Equ:5}
\end{eqnarray}
We can then set the bound on the parameter $\zeta_{AZ}$:
\begin{eqnarray}
- 0.0077 > \zeta _ { A Z } > 0.0037.
\label{Equ:6}
\end{eqnarray}

\section{Further Study}
We are planning to improve the analysis by adding the $h \xrightarrow{} WW^*$ channel. The branching ratio of this channel is around 21$\%$ corresponding to about 50 event for 2000 fb$^{-1}$. The main background we expect in this channel would be $e^+e^-\to W^+W^-$ with a hard ISR photon, while in the $h\to b\bar{b}$ channel, the main background is $e^+e^-\to b\bar{b}$ with a hard ISR photon, which is significantly enhanced due to the radiative return to $Z$-pole. We would hence expect a higher signal to background ratio in the $h \to WW^*$ channel.

After $h\to WW^*$ channel is completed, the experimental bound on $\zeta_{AZ}$ will be translated into a bound on Dimension-6 operators.

%・Understand the role of this measurement in one global EFT analysis

\section*{Acknowledgements}
We would like to thank the LCC generator working group and the ILD software working group for providing the simulation and reconstruction tools and producing the Monte Carlo samples used in this study.
This work has benefited from computing services provided by the ILC Virtual Organization, supported by the national resource providers of the EGI Federation and the Open Science GRID.

\end{document}